\documentclass[referee]{raa}
\usepackage{graphicx,times}
\usepackage{longtable,lscape}
\begin{document}

   \title{The electromagnetic and gravitational-wave radiations of X-ray transient CDF-S XT2}

   \volnopage{Vol.0 (200x) No.0, 000--000}      %%preserved for Editor. DOn't remove!
   \setcounter{page}{1}          %%starting page, preserved for Editor. DOn't remove!

   \author{Houjun L\"{u}
      \inst{1\dag}
   \and Yong Yuan
      \inst{1}
   \and Lin Lan
      \inst{1}
   \and Binbin Zhang
      \inst{2,3}
   \and Jinhang Zou
      \inst{4}
   \and En-Wei Liang
	 \inst{1}
   }

   \institute{Guangxi Key Laboratory for Relativistic Astrophysics,
School of Physical Science and Technology, Guangxi University, Nanning 530004, China; {\it
             lhj@gxu.edu.cn}\\
        \and
             School of Astronomy and Space Science, Nanjing University, Nanjing 210093, China; \\
        \and
             Key Laboratory of Modern
Astronomy and Astrophysics (Nanjing University), Ministry of Education, Nanjing 210093, China;\\
        \and
             Space Science and Technology, Hebei normal university, Shijiazhuang 050000, China.\\
   }

   \date{Received~~201X month day; accepted~~201X~~month day}

\abstract{Binary neutron star (NS) mergers may result in remnants of supra-massive or even stable
NS, which have been supported indirectly by observed X-ray plateau of some gamma-ray bursts (GRBs)
afterglow. Recently, Xue et al. (2019) discovered a X-ray transient CDF-S XT2 that is powered by a
magnetar from merger of double NS via X-ray plateau and following stepper phase. However, the decay
slope after the plateau emission is a little bit larger than the theoretical value of spin-down in
electromagnetic (EM) dominated by losing its rotation energy. In this paper, we assume that the
feature of X-ray emission is caused by a supra-massive magnetar central engine for surviving
thousands of seconds to collapse black hole. Within this scenario, we present the comparisons of
the X-ray plateau luminosity, break time, and the parameters of magnetar between CDF-S XT2 and
other short GRBs with internal plateau samples. By adopting the collapse time to constrain the
equation of state (EOS), we find that three EOSs (GM1, DD2, and DDME2) are consistent with the
observational data. On the other hand, if the most released rotation energy of magnetar is
dominated by GW radiation, we also constrain the upper limit of ellipticity of NS for given EOS,
and it is range in $[0.32-1.3]\times 10^{-3}$. Its GW signal can not be detected by aLIGO or even
for more sensitive Einstein Telescope in the future.  }

   \authorrunning{L\"{u} et al. }            %author_head in even pages
   \titlerunning{EM and GW radiations of X-ray transient CDF-S XT2}  % title_head in odd pages
   \maketitle
%% The author head (on even pages) and the title head (on odd pages) will be
%% automatically extracted from \author{} and \title{}. Whenever the title is too long,
%% you will be asked to supply a shorter one by inserting either \authorrunning{} or
%% \titlerunning{} before \maketitle. Anyway, you can specify your own heads.
%%
%%
%% Note: In the following text body of your manuscript, please note several differences from
%%       other major journals:
%% (1) \subsection{Please Capitalize the First Letter of Each Notional Word in Subsection Title}
%% (2) Please Capitalize the First Letter of Each Notional Word in all tables' captions

%
%________________________________________________ sections below
%

%%%%%%%%%%%%%%%%%%%%%%%%%%%%%%%%%%%%%%%%%%%%%%%%%%%%%%%%%%%%%%%%
\section{Introduction}           %% first-level sections will be auto-capitalized
\label{sect:intro}

The merger of a binary neutron star (NS) system is thought to be potential sources of producing
both gravitational wave (GW) and associated electromagnetic (EM) signals (Berger 2014 for review).
One solid case of producing GW signal and associated EM (GW170817 and GRB 170817A, as well as
kilonova AT2017gfo), is already detected by Advanced LIGO, VIRGO, and other telescopes (Abbott et
al. 2017; Goldstein et al. 2017; Coulter et al. 2017; Zhang et al. 2018). However, the remnants of
double NS merger remain an open question.

From the theoretical point of view, there are four different types of merger remnants that are
dependent on the total mass of the system and the poorly known NS equation of state (EOS; Rosswog
et al. 2000; Dai et al. 2006; Fan \& Xu 2006; Metzger et al. 2010; Rezzolla et al. 2011; Giacomazzo
\& Perna 2013; Zhang 2013; Lasky et al. 2014; Li et al. 2016). (1) a promptly formed BH (Hotokezaka
et al. 2011); (2) a hyper-massive NS, can be survived for $\sim100$ ms before collapsing into a BH
(Baumgarte et al. 2000; Shibata \& Taniguchi 2006; Palenzuela et al. 2015); (3) a supra-massive NS,
is supported by rigid rotation and survives for seconds to hours (Dai et al. 2006; Rowlinson et al.
2010; Hotokezaka et al. 2013; Zhang 2014; L\"{u} et al. 2015; Gao et al. 2016; Kiuchi et al. 2018);
(4) a stable NS (Dai \& Lu 1998; Zhang \& M\'esz\'aros 2001; Yu et al. 2010; Metzger et al. 2011;
Bucciantini et al. 2012; L\"{u} \& Zhang 2014).

Recently, Xue et al. (2019) discovered a X-ray transient CDF-S XT2 that is associated with a galaxy
at redshift $z=0.738$, and its X-ray light curve is consistent with magnetar central engine model
which is originated from double neutron star merger. The magnetar parameters are inferred by
invoking the its X-ray plateau and followed decay segment in their work, and found that they are
consistent with the parameters of magnetar in typical short GRBs. However, the decay slope after
the plateau emission is a little bit larger than the theoretical value of magnetar spin-down. On
the other hand, a lower efficiency ($\eta=0.001$) is adopted to estimate the parameters of magnetar
for this transient in Xue et al. (2019). Here, we assume that the feature of X-ray emission is
caused by a supra-massive magnetar central engine for surviving thousands of seconds to collapse
black hole. Within this scenario, most rotation energy of magnetar may be dissipated in two ways.
One is that the most of rotation energy is transformed into kinetic energy with injecting pulsar
wind (Xiao \& Dai 2019). The other possibility is that most of the rotational energy was carried
away via the strong gravitational wave radiation (Fan et al. 2013; Lasky \& Glampedakis 2016;
L\"{u} et al. 2018). In this paper, by considering two scenarios of rotation energy loss of
magnetar for post-merger (i.e., EM dominated or GW dominated), we infer the surface magnetic field
and initial period of NS, and constrain the EOS and ellipticity of NS, as well as detection
probability of GW.

This paper is organized as follows. The empirical fitting of X-ray light curve for transient CDF-S
XT2 is presented in section 2. Some comparisons between CDF-S XT2 and other short GRBs with
internal plateau, as well as EOS are shown in section 3. In section 4, we constrain the ellipticity
of NS, and calculate the detection probability of GW. The conclusions, along with some discussions,
are presented in Section 5. Throughout this paper, a concordance cosmology with parameters $H_{\rm
0} = 71\rm~km~s^{-1}~Mpc^{-1}$, $\Omega_M=0.30$, and $\Omega_{\Lambda}=0.70$ is adopted.

\section{Light curve fit and central engine of CDF-S XT2}
\subsection{Light curve fit of CDF-S XT2}
\begin{figure*}
\includegraphics[angle=0,scale=0.5]{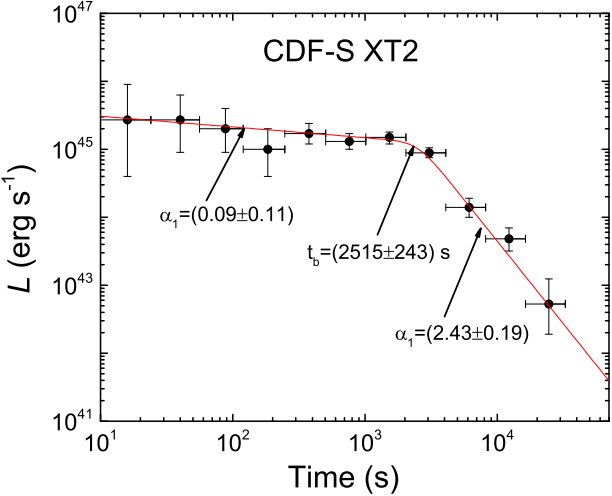}
\centering \caption{X-ray light curve of CDF-S XT2. The red solid line is the fit with smooth
broken power-law model.} \label{fig:LC}
\end{figure*}

The X-ray data of CDF-S XT2 observed by Chandra within energy band 0.3-10 keV are taken from Xue et
al. (2019). We perform a temporal fit to the light curve with a smooth broken power law model,
which is expression as
\begin{eqnarray}
L = L_{0} \left[\left(\frac{t}{t_b}\right)^{\omega\alpha_1}+
\left(\frac{t}{t_b}\right)^{\omega\alpha_2}\right]^{-1/\omega}
\end{eqnarray}
where $t_{b}(2525\pm 242)$ s is the break time, $L_b=L_0 \cdot 2^{-1/\omega}=(1.28\pm 0.16)\times
10^{45}\rm ~erg~s^{-1}$ is the luminosity at the break time $t_b$, $\alpha_1=(0.09\pm0.11)$ and
$\alpha_2=(2.43\pm0.19)$ are decay indices before and after the break, respectively. The $\omega$
describes the sharpness of the break. The larger the $\omega$ parameter, the sharper the break, and
$\omega=3$ is fixed for the light curve fitting. An IDL routine named ``mpfitfun.pro'' is employed
for our fitting (Markwardt 2009). This routine performs a Levenberg-Marquardt least-square fit to
the data for a given model to optimize the model parameters. The light curve fit is shown in Figure
\ref{fig:LC}.

\subsection{Central engine of CDF-S XT2}
X-ray transient CDF-S XT2 associated with a galaxy at redshift $z=0.738$, lies in the outskirts of
its star-forming host galaxy with a moderate offset from the galaxy center, and no significant
source-like gamma-ray emission signal above background. Those observed properties are similar with
other typical short GRBs, but in off-axis observed (Xue et al. 2019). On the other hand, the
estimated event rate density of this event is similar with double NS merger rate density inferred
from the detection of GW170817, suggesting that the progenitor of this event is likely from double
NS merger (Xue et al. 2019). Moreover, the observed X-ray plateau of CDF-S XT2 is consistent with
wind dissipation of magnetar central engine, and it indicate that the remnants of such double NS
merger should be either supra-massive NS or stable NS. However, the decay slope after the plateau
emission ($t^{-2.43}$) is a little bit larger than the theoretical value of spin-down ($t^{-2}$) in
electromagnetic (EM) dominated by losing its rotation energy. Here, we assume that the feature of
X-ray emission is caused by a supra-massive magnetar central engine for surviving thousands of
seconds to collapse black hole.

\begin{figure*}
\includegraphics[angle=0,scale=0.5]{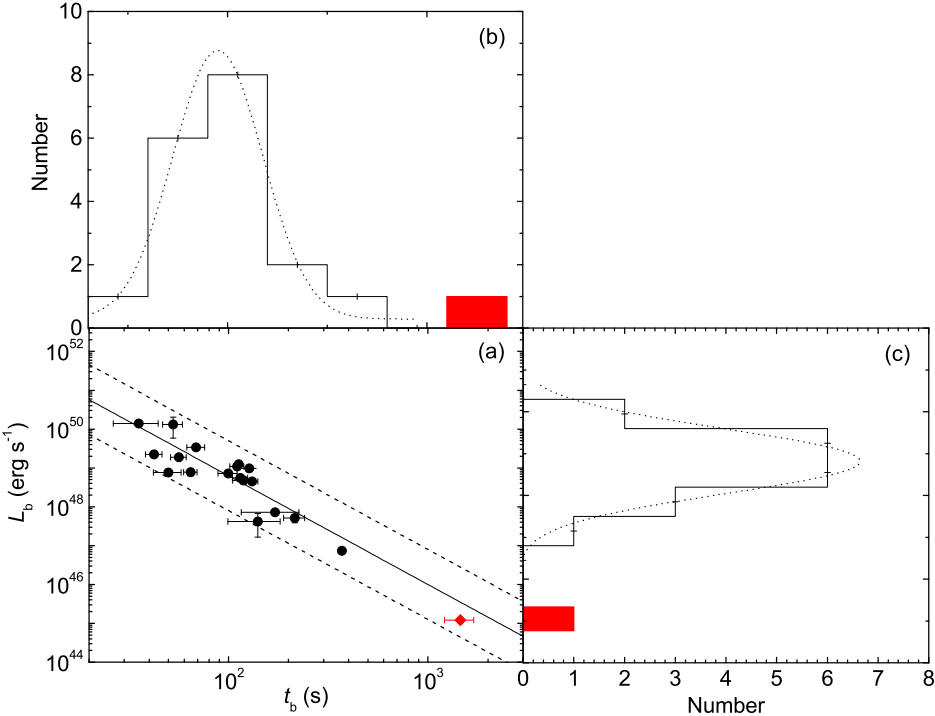}
\centering \caption{(a): X-ray plateau luminosity ($L_0$) as function of collapse time ($t_{\rm
col}$) for short GRBs with internal plateau (black dots) and X-ray transient CDF-S XT2 (red
diamond). The black solid line is the the best fit with power-law model, and the two dashed lines
mark the 2$\sigma$ region of the correlation, respectively. (b) and (c): Distributions of $t_{\rm
col}$ and $L_0$ with best-fit Gaussian profiles, respectively.} \label{fig:Ltao}
\end{figure*}

In order to compare the properties of CDF-S XT2 with other short GRBs with internal plateau, Fig.
\ref{fig:Ltao} shows the correlation between break luminosity ($L_b$) and collapse time ($\tau_{\rm
col}=t_{\rm b}/(1+z)$), as well as the distributions of $L_b$ and $\tau_{\rm col}$. We find that
the CDF-S XT2 fall into the 2$\sigma$ deviation in $L_b-\tau_{\rm col}$ diagram, suggesting that
the other short GRBs with internal plateau samples shared similar central engine with the CDF-S
XT2. However, the distributions of luminosity and collapse time of the CDF-S XT2 is much lower and
longer than other short GRBs with internal plateau samples, respectively. It may be caused by the
directions of observations (i.e., on and off-axis with short GRBs and the CDF-S XT2), or having
different populations of magnetar.

If we believe a supramassive NS as a potential candidate central engine of CDF-S XT2. One
interesting question is that what is the energy loss channel of the rotating magnetar, dominated by
magnetic dipole or GW radiation. We will discuss more details for the rotation energy loss of
magnetar dominated by EM or GW radiations.

\section{The rotation energy loss of magnetar via EM emission}
\subsection{The derived parameters of magnetar}
The energy reservoir of a millisecond magnetar is the total rotation energy, which reads as
\begin{eqnarray}
E_{\rm rot} = \frac{1}{2} I \Omega^{2}
\simeq 2 \times 10^{52}~{\rm erg}~
M_{1.4} R_6^2 P_{-3}^{-2},
\label{Erot}
\end{eqnarray}
where $I$ is the moment of inertia, $\Omega$, $P$, $R$, and $M$ are the angular frequency, rotating
period, radius, and mass of the neutron star, respectively. The convention $Q = 10^x Q_x$ in cgs
units is adopted. A magnetar spinning down loses its rotational energy via both magnetic dipole
torques ($L_{\rm EM}$) and GW ($L_{\rm GW}$) radiations (Zhang \& M{\'e}sz{\'a}ros 2001; Fan et al.
2013; Giacomazzo \& Perna 2013; Lasky \& Glampedakis 2016; L\"{u} et al. 2018),
\begin{eqnarray}
-\frac{dE_{\rm rot}}{dt} = -I\Omega \dot{\Omega} &=& L_{\rm total} = L_{\rm EM} + L_{\rm GW} \nonumber \\
&=& \frac{B^2_{\rm p}R^{6}\Omega^{4}}{6c^{3}}+\frac{32GI^{2}\epsilon^{2}\Omega^{6}}{5c^{5}},
\label{Spindown}
\end{eqnarray}
where $B_{\rm p}$ is the surface magnetic field at the pole and
$\epsilon=2(I_{xx}-I_{yy})/(I_{xx}+I_{yy})$ is the ellipticity describing how large of the neutron
star deformation. $\dot{\Omega}$ is the time derivative of the angular frequency. One can find that
for a magnetar with given R and I, its $L_{\rm EM}$ depends on $B_{\rm p}$ and $\Omega$, and
$L_{\rm GW}$ depends on $\epsilon$ and $\Omega$.

If the rotation energy loss of magnetar is dominated by EM emission, one has
\begin{eqnarray}
L_{\rm EM}\simeq -I\Omega \dot{\Omega}=\frac{\eta B^2_{\rm p}R^{6}\Omega^{4}}{6c^{3}}.
\label{EM_dominated}
\end{eqnarray}
whre $\eta$ is the efficiency of converting the magnetar wind energy into X-ray radiation. The
characteristic spin-down luminosity ($L_{\rm EM, sd}$) and time scale ($\tau_{\rm EM, sd}$) of
magnetar can be given as,
\begin{eqnarray}
L_{\rm EM,sd}&=&\frac{\eta B^2_{p}R^6\Omega^{4}_{0}}{6c^3} \nonumber \\
&\simeq&1.0 \times 10^{46}~{\rm erg~s^{-1}} (\eta_{-3}B_{p,15}^2 P_{0,-3}^{-4} R_6^6),
\label{spinlu_em}
\end{eqnarray}
\begin{eqnarray}
\tau_{\rm EM,sd}&=&\frac{3c^{3}I}{B_{p}^{2}R^{6}\Omega_{0}^{2}} \nonumber \\
&\simeq&2.05 \times 10^3~{\rm s}~ (I_{45} B_{p,15}^{-2} P_{0,-3}^2 R_6^{-6}),
\label{spintau_em}
\end{eqnarray}
where $\Omega_0$ and $P_0$ are initial angular frequency and period at $t=0$, respectively.

\begin{figure*}
\includegraphics[angle=0,scale=0.6]{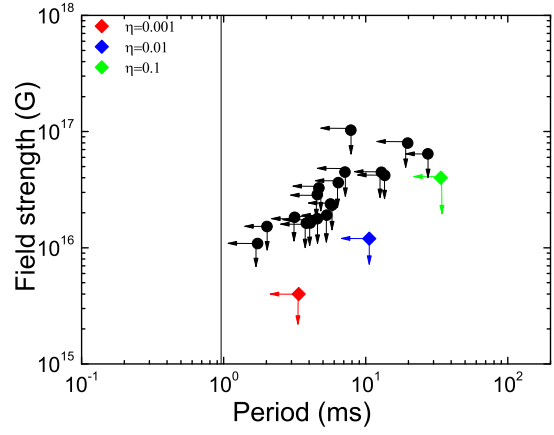}
\centering\caption{Inferred magnetar parameters, initial spin period $P_0$ vs. surface polar cap
magnetic field strength $B_{\rm p}$ derived for short GRBs with internal plateau (black dots) and
X-ray transient CDF-S XT2 (diamond) with $\eta=0.1, 0.01$ and 0.001. The vertical solid line is the
break-up spin period limit for a neutron star (Lattimer \& Prakash 2004).} \label{fig:BpP0}
\end{figure*}

Within the magnetar central engine scenario, the observed plateau luminosity is closed to $L_{\rm
b}$, which is roughly equal to $L_{\rm EM,sd}$, and $\tau_{\rm EM,sd}>\tau_{\rm col}$. One can
derive the magnetar parameters $B_{\rm p}$ and $P_0$,
\begin{eqnarray}
B_{\rm p,15} = 2.05(\eta^{1/2}_{-3}I_{45} R_6^{-3} L_{\rm EM, sd, 46}^{-1/2} \tau_{\rm EM, sd, 3}^{-1})~\rm
G,
\label{Bp}
\end{eqnarray}
\begin{eqnarray}
P_{0,-3} = 1.42(\eta^{1/2}_{-3}I_{45}^{1/2} L_{\rm EM, sd, 46}^{-1/2} \tau_{\rm EM, sd, 3}^{-1/2})~\rm s.
\label{P0}
\end{eqnarray}
As radiation efficiency $\eta$ depends strongly on the injected luminosity and wind saturation
Lorentz factor (Xiao\& Dai 2019). By adopting the lower limit of $\tau_{\rm EM,sd}$, we derive the
upper limits of $P_0$ and $B_{\rm p}$ with different $\eta$ values. One has $P_0 < 3.4 \times
10^{-3}$ s and $B_{\rm p}< 4 \times 10^{15}$ G for $\eta=0.001$, $P_0 < 10.6 \times 10^{-3}$ s and
$B_{\rm p}< 1.2 \times 10^{16}$ G for $\eta=0.01$, and $P_0 < 33.8 \times 10^{-3}$ s and $B_{\rm
p}< 4 \times 10^{16}$ G for $\eta=0.1$. Figure \ref{fig:BpP0} shows the $B_{\rm p}-P_0$ diagram for
X-ray transient CDF-S XT2 with different $\eta$ values, and compares with other short GRBs with
internal plateau sample taken from L\"{u} et al. (2015). It seems that small $P_0$ required by
supra-massive magnetar is needed lower radiation efficiency, and estimated $B_{\rm p}$ of CDF-S XT2
is lower than other typical short GRBs samples for smaller $P_0$. It may be either off-axis
observations or different population of CDF-S XT2 by comparing with short GRBs.

\subsection{Equation of state of NS}
The inferred collapsing time can be used to constrain the neutron star EOS (Lasky et al. 2014; Ravi
\& Lasky 2014; L\"{u} et al. 2015). The basic formalism is as follows.

The standard dipole spin-down formula gives (Shapiro \& Teukolsky 1983)
\begin{eqnarray}
P(t) &=& P_{0} (1+\frac{4\pi ^{2}}{3c^{3}}\frac{B_{p}^{2}R^{6}}{I P_{0}^{2}}t)^{1/2}\nonumber \\
&=&P_{0} (1+\frac{t}{\tau_{\rm EM, sd}})^{1/2}.
\label{Pt}
\end{eqnarray}
The maximum NS mass for a non-rotating NS ($M_{\rm TOV}$) can be derived for given EOS of NS. The
maximum gravitational mass ($M_{\rm max}$) depends on spin period, read as (Lyford et al. 2003)
\begin{eqnarray}
M_{\rm max} = M_{\rm TOV}(1+\hat{\alpha} P^{\hat{\beta}})
\label{Mt1}
\end{eqnarray}
where $\hat{\alpha}$, $\hat{\beta}$, and $M_{\rm TOV}$ depend on the EOS of NS.

As the neutron star spins down, the centrifugal force can no longer sustain the star, and the NS
will collapse into a black hole. By using Equation(\ref{Pt}) and (\ref{Mt1}), one can derive the
collapse time as function of $M_{\rm p}$,
\begin{eqnarray}
t_{\rm col} &=& \frac{3c^{3}I}{4\pi^{2}B_{\rm p}^{2}R^{6}}[(\frac{M_{\rm p}-M_{\rm
TOV}}{\hat{\alpha} M_{\rm
TOV}})^{2/\hat{\beta}}-P_{0}^{2}]\nonumber \\
&=&\frac{\tau_{\rm EM, sd}}{P_{\rm 0}^{2}}[(\frac{M_{\rm p}-M_{\rm TOV}}{\hat{\alpha} M_{\rm
TOV}})^{2/\hat{\beta}}-P_{0}^{2}].
\label{tcol}
\end{eqnarray}
Here, we consider 12 EOS that are reported in the literatures (Lasky et al. 2014; Ravi \& Lasky
2014; Li et al. 2016; Ai et al. 2018). The basic parameters of those EOS are shown in Table 1.

\begin{figure*}
\includegraphics[angle=0,scale=0.5]{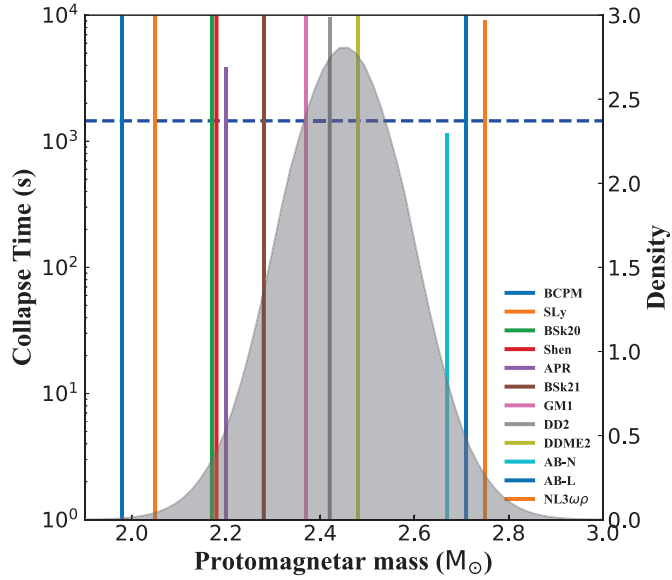}
\centering\caption{Collapse time as a function of the protomagnetar mass of CDF-S XT2 for different
EOS (color lines). The shaded region is the protomagnetar mass distribution derived from the total
mass distribution of the Galactic NS¨CNS binary systems. The horizontal dashed line is the collapse
time in the rest frame.} \label{fig:EOS}
\end{figure*}

As noted, one can infer $B_p$, $P_0$, and $t_{\rm col}$ from the observations by adopting
$\eta=0.001$. Following the method of Lasky et al. (2014), a tight mass distribution of the our
Galactic binary NS population is adopted (e.g., Valentim et al. 2011; Kiziltan et al. 2013), and
one can infer the expected distribution of proto-magnetar masses, which is found to be
$M_p=2.46^{0.13}_{-0.15}M_\odot$. For X-ray transient CDF-S XT2, the lower limit of $\tau_{\rm EM,
sd}=t_{\rm col}$ is derived. Figure \ref{fig:EOS} presents the collapse time ($t_{\rm col}$) as a
function of protomagnetar mass ($M_p$) for CDF-S XT2 with different EOS. Our results show that the
GM1, DD2, and DDME2 models give an $M_{\rm p}$ band falling within the 2$\sigma$ region of the
protomagnetar mass distribution, so that the correct EOS should be close to those three models. The
maximum mass for non-rotating NS in those three models are $M_{\rm TOV}=2.37M_{\odot}$,
$2.42M_{\odot}$, and $2.48M_{\odot}$, respectively.

\section{The rotation energy loss of magnetar via GW radiation}
A survived supra-massive NS central engine requires a more fast spinning ($P_0\sim 1$ ms) to
support the gravitational force (Fan et al. 2013; Gao et al. 2013; Yu et al. 2013; Zhang 2013; Ho
2016; Lasky \& Glampedakis 2016). As mentioned above, the estimated periods of magnetar are
considerably longer ($\eta=0.01$ and 0.1) than that expected in the double neutron star merger
model. It seems that $\eta$ should be as low as 0.001 or even smaller to obtain the lower period of
magnetar. If this is the case, the rotation energy loss of magnetar is either transformed to
kinetic energy of outflow or dominated by GW radiation (Lan et al. 2020). Xiao \& Dai (2019)
present a more details for the first situation. In this section, we focus on considering the most
rotation energy of magnetar dissipated via GW radiation.

\subsection{Constraining the ellipticity of NS}
Within GW dominated scenario, one has (L\"{u} et al. 2018),
\begin{eqnarray}
L_{\rm GW}\simeq -I\Omega \dot{\Omega}=\frac{32GI^{2}\epsilon^{2}\Omega^{6}}{5c^{5}}.
\label{GW_dominated}
\end{eqnarray}
The characteristic spin-down luminosity ($L_{\rm GW, sd}$) and time scale ($\tau_{\rm GW, sd}$) of
NS can be given as,
\begin{eqnarray}
L_{\rm GW,sd}&=& \frac{32GI^{2}\epsilon^{2}\Omega_0^{6}}{5c^{5}} \nonumber \\
&\simeq&1.08 \times 10^{48}~{\rm erg~s^{-1}}(I_{45}^2 \epsilon_{-3}^{2} P_{0,-3}^{-6}),
\label{spinlu_gw}
\end{eqnarray}
\begin{eqnarray}
\tau_{\rm GW,sd}&=&\frac{5c^{5}}{128GI\epsilon^2\Omega^4_0} \nonumber \\
&\simeq&9.1 \times 10^3~{\rm s}~ (I^{-1}_{45}\epsilon_{-3}^{-2} P_{0,-3}^4 ).
\label{spintau_gw}
\end{eqnarray}
The supra-massive NS of CDF-S XT2 has collapse into black hole before it is spin-down, so that one
has $\tau_{\rm GW,sd}>\tau_{\rm col}$. Combining with Equation (\ref{spintau_gw}), the upper limit
of ellipticity ($\epsilon$) can be expressed as
\begin{eqnarray}
\epsilon<2.5\times 10^{-3}I_{45}^{-1/2}P_{0,-3}^{2}.
\label{ellipticity}
\end{eqnarray}
The maximum value of $\epsilon$ for different EOS with $P_0=1$ ms are shown in Table 1. We find
that those values are in the range of $[0.32-1.3]\times 10^{-3}$. This upper limit value is larger
than the maximum elastic quadrupole deformation of conventional neutron stars, but is comparable to
the upper limit derived for crystalline colour-superconducting quark matter (Lin 2007;
Johnson-McDaniel \& Owen 2014).

\subsection{Detection Probability of a GW}
\begin{figure*}
\includegraphics[angle=0,scale=0.6]{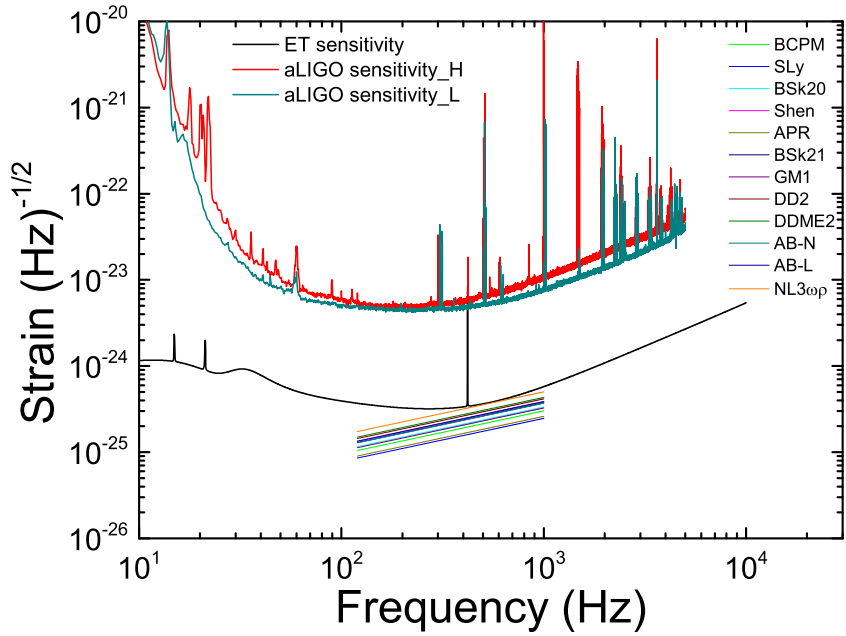}
\centering\caption{Gravitational-wave strain evolution with frequency for CDF-S XT2 with different
EOS at distances $D_{\rm L}=4480$ Mpc (color lines). The black solid line is the sensitivity limits
for ET, and the red and dark cyan solid lines are are the sensitivity limits for aLIGO-Hanford and
aLIGO-Livingston, respectively. The data of noise curve are taken from the website: $\rm
https://git.ligo.org/lscsoft/bilby/-/tree/master/bilby/gw/detector/noise\_curves$}
\label{fig:LIGOET}
\end{figure*}

If most of the rotation energy is released via GW radiation with a frequency $f$, the GW strain for
a rotating neutron star at distance $D_{\rm L}$ can be expressed as,
\begin{eqnarray}
h(t)=\frac{4G I \epsilon}{D_{\rm L}c^{4}} \Omega(t)^{2}
\end{eqnarray}
The signal-to-noise ratio of optimal matched filter can be expressed as
\begin{eqnarray}
\rho^2=\int^{f_2}_{f_1}\frac{\tilde{f}^2(f)}{S_{h}(f)}df
\end{eqnarray}
where $f_1$ and $f_2$ are the initial and final GW frequencies, respectively. $\tilde{h}(f)$ is the
Fourier transform of $h(t)$, namely $\tilde{h}(f)=h(t)\sqrt{dt/df}$. $S_{h}(f)$ is the noise power
spectral density of the detector (Lasky \& Glampedakis 2016). The characteristic amplitude of GW
from a rotating NS can be estimated as (Corsi \& M\'{e}sz\'{a}ros 2009; Hild et al. 2011; Lasky \&
Glampedakis 2016; L\"{u} et al. 2017),
\begin{eqnarray}
h_{\rm c}& = & f h(t)\sqrt{\frac{dt}{df}}=\frac{f}{D_{\rm L}}\sqrt{\frac{5GI}{2c^{3}f}} \nonumber \\
&\approx& 8.22\times 10^{-24} \biggl(\frac{I}{10^{45}~\rm g\,cm^{2}}\frac{f}{1~\rm
kHz}\biggr)^{\!1/2}\biggl(\frac{D_{\rm
L}}{100~\rm Mpc}\biggr)^{\!-1}~\rm.
\nonumber
\\
\end{eqnarray}
For X-ray transient CDF-S XT2, its redshift $z=0.738$ corresponds to $D_{\rm L}\sim 4480~\rm Mpc$.
By adopting the frequency range of GW from $f=120~\rm Hz$ to $1000~\rm Hz$, one can estimate the
maximum value of the strain $h_{\rm c}$ for different EOS of NS. The estimated values of $h_{\rm
c}$ are reported in Table 1. The maximum value of the strain $h_{\rm c}$ for NL3$\omega\rho$ is
about $5\times 10^{-25}$, which is about one order of magnitude smaller than the advanced-LIGO
sensitivity, and also less than more sensitive Einstein Telescope (ET; see Figure
\ref{fig:LIGOET}). It means that even if the merger remnant of double NS of this transient is a
millisecond massive NS, the post merger GW signal is undetectable when the rotation energy of the
NS is taken away by the GW radiation.

\section{Conclusions and Discussion}
The X-ray transient CDF-S XT2 associated with a galaxy at redshift $z=0.738$, lies in the outskirts
of its star-forming host galaxy with a moderate offset from the galaxy center, and no significant
source-like gamma-ray emission signal above background (Xue et al. 2019). Moreover, the estimated
event rate density of this event is similar with double NS merger rate density inferred from the
detection of GW170817, and the observed X-ray plateau is consistent with wind dissipation of
magnetar central engine. Those observed evidences support that the progenitor of this event is
likely from double NS merger, and the remnants of such double NS merger should be either
supra-massive NS or stable NS. Moreover, Xiao et al (2019) proposed that both the light curve and
spectral evolution of CDF-S XT2 can be well explained by the internal gradual magnetic dissipation
process in an ultra-relativistic wind. Sun et al. (2109) also presented that this transient is only
observed from different zone, defined as free zone where the X-ray emission from magnetar spin-down
can escape freely. Alternatively, Peng et al. (2019) argued that this transient is possible from
tidal disruption event.

The decay slope after the plateau emission of CDF-S XT2 ($t^{-2.43}$) is a little bit larger than
the theoretical value of spin-down ($t^{-2}$)in electromagnetic (EM) dominated by losing its
rotation energy. In this work, we assume that the feature of X-ray emission is caused by a
supra-massive magnetar central engine for surviving thousands of seconds to collapse black hole.
Within this scenario, in order to compare the observed properties of X-ray emission between CDF-S
XT2 and other short GRBs with internal plateau, we show the correlation between break luminosity
and collapse time, as well as the distributions of them. We find that the CDF-S XT2 fall into the
2$\sigma$ deviation in $L_b-\tau_{\rm col}$ diagram, suggesting that the other short GRBs with
internal plateau samples shared similar central engine with the CDF-S XT2. However, the
distributions of luminosity and collapse time of the CDF-S XT2 is much lower and longer than other
short GRBs with internal plateau samples, respectively. It may be caused by the directions of
observations (i.e., on- and off-axis with short GRBs and the CDF-S XT2), or having different
populations of magnetar.

On the other hand, one consider two channels of rotation energy loss of supra-massive magnetar, one
is EM dominated, and the other is GW dominated. Within the first scenario, we estimate the
parameters of magnetar (i.e., $B_{\rm p}$ and $P_0$) for given different radiation efficiency, as
well as constraining the EOS of NS. It seems that small $P_0$ required by supra-massive magnetar is
needed lower radiation efficiency, and estimated $B_{\rm p}$ of CDF-S XT2 is lower than other
typical short GRBs samples for smaller $P_0$. Moreover, we find that three EOSs (GM1, DD2, and
DDME2) are consistent with the observational data of CDF-S XT2. Within the second scenario, we
constrain the upper limit of ellipticity of NS for given different EOS, it is range of
$[0.32-1.3]\times 10^{-3}$. By calculating the possible GW radiation for different EOS, we find
that its GW radiation can not be detected by aLIGO or even for more sensitive Einstein Telescope in
the future.

\begin{acknowledgements}
We acknowledge the use of public data from the {\em Swift} and {\em Fermi} data archive, and the UK
{\em Swift} Science Data Center. We thank the anonymous referee for helpful comments. This work is
supported by the National Natural Science Foundation of China (grant Nos.11922301, 11851304,
11533003, and 11833003), the Guangxi Science Foundation (Grant Nos. 2017GXNSFFA198008,
2018GXNSFGA281007, and AD17129006). The One-Hundred-Talents Program of Guangxi colleges, Bagui
Young Scholars Program (LHJ), and special funding for Guangxi distinguished professors (Bagui
Yingcai \& Bagui Xuezhe). BBZ acknowledges support from a national program for young scholars in
China, Program for Innovative Talents and Entrepreneur in Jiangsu, and a National Key Research and
Development Programs of of China (2018YFA0404204).
\end{acknowledgements}

%&&&&&&&&&&&&&&&&&&&&&&&&&&&&&&&&&&&&&&&&&&&&&&&&&&&&&&&&&&&&&&&&&&&&&&&&&&&&&&&&&&

%********************************Table1*****************************************************
\clearpage
\begin{table*}
\caption{The basic parameters of EOS of NS} \label{table:1} \centering
\begin{tabular}{c c c c c c c c}
\hline\hline
&$M_{TOV}$ & R & $I$ & $\hat{\alpha}$ & $\hat{\beta}$ & $\epsilon$ & $h_{\rm
c}(f)$ \\
&$(M_{\odot}$) & (km) & $(10^{45}~{\rm
g~cm^{2}})$ & $(10^{-10}~{s^{-\hat{\beta}}})$ &  & $(10^{-3})$ & $(10^{-25})$ \\
\hline
BCPM             &1.98  &9.94   &2.86  &3.39  &-2.65  &$1.5$   &$3.02$  \\
SLy              &2.05  &9.99   &1.91  &1.60  &-2.75  &$1.8$   &$2.47$  \\
BSk20            &2.17  &10.17  &3.50  &3.39  &-2.68  &$1.3$   &$3.34$  \\
Shen             &2.18  &12.40  &4.68  &4.69  &-2.74  &$1.2$   &$3.87$  \\
APR              &2.20  &10.0   &2.13  &0.303 &-2.95  &$1.7$   &$2.61$  \\
BSk21            &2.28  &11.08  &4.37  &2.81  &-2.75  &$1.2$   &$3.74$  \\
GM1              &2.37  &12.05  &3.33  &1.58  &-2.84  &$1.4$   &$3.26$  \\
DD2              &2.42  &11.89  &5.43  &1.37  &-2.88  &$1.1$   &$4.16$  \\
DDME2            &2.48  &12.09  &5.85  &1.966 &-2.84  &$1.0$   &$4.32$  \\
AB-N             &2.67  &12.90  &4.30  &0.112 &-3.22  &$1.2$   &$3.71$  \\
AB-L             &2.71  &13.70  &4.70  &2.92  &-2.82  &$1.2$   &$3.87$  \\
NL3$\omega\rho$  &2.75  &12.99  &7.89  &1.706 &-2.88  &$0.89$  &$5.02$  \\
\hline
\end{tabular}
\end{table*}
%_____________________________________________________________

%**************************Figures**********************************************************

%&&&&&&&&&&&&&&&&&&&&&&&&&&&&&&&&&&&&Figures&&&&&&&&&&&&&&&&&&&&&&&&&&&&&&&&&&&&&&&&&&&&&&&

\clearpage

\begin{thebibliography}{999}
\bibitem[Abbott et al.(2017)]{2017ApJ...848L..13A} Abbott, B.~P., Abbott, R., Abbott, T.~D., et
    al.\ 2017, \apjl, 848, L13
\bibitem[Ai et al.(2018)]{2018ApJ...860...57A} Ai, S., Gao, H., Dai, Z.-G., et al.\ 2018, \apj,
    860, 57
\bibitem[Baumgarte et al.(2000)]{2000ApJ...528L..29B} Baumgarte, T.~W., Shapiro, S.~L., \& Shibata,
    M.\ 2000, \apjl, 528, L29
\bibitem[Berger(2014)]{2014ARA&A..52...43B} Berger, E.\ 2014, \araa, 52, 43
\bibitem[Bucciantini et al.(2012)]{2012MNRAS.419.1537B} Bucciantini, N., Metzger, B.~D., Thompson,
    T.~A., \& Quataert, E.\ 2012, \mnras, 419, 1537
\bibitem[Corsi \& M{\'e}sz{\'a}ros(2009)]{2009ApJ...702.1171C} Corsi, A., \& M{\'e}sz{\'a}ros, P.\
    2009, \apj, 702, 1171
\bibitem[Coulter et al.(2017)]{2017Sci...358.1556C} Coulter, D.~A., Foley, R.~J., Kilpatrick,
    C.~D., et al.\ 2017, Science, 358, 1556
\bibitem[Dai \& Lu(1998)]{1998A&A...333L..87D} Dai, Z.~G., \& Lu, T.\ 1998, \aap, 333, L87
\bibitem[Dai et al.(2006)]{2006Sci...311.1127D} Dai, Z.~G., Wang, X.~Y., Wu, X.~F., \& Zhang, B.\
    2006, Science, 311, 1127
\bibitem[Fan et al.(2013)]{2013PhRvD..88f7304F} Fan, Y.-Z., Wu, X.-F., \& Wei, D.-M.\ 2013, \prd,
    88, 067304
\bibitem[Fan \& Xu(2006)]{2006MNRAS.372L..19F} Fan, Y.-Z., \& Xu, D.\ 2006, \mnras, 372, L19
\bibitem[Gao et al.(2013)]{2013ApJ...771...86G} Gao, H., Ding, X., Wu, X.-F., Zhang, B., \& Dai,
    Z.-G.\ 2013, \apj, 771, 86
\bibitem[Gao et al.(2016)]{2016PhRvD..93d4065G} Gao, H., Zhang, B., \& L{\"u}, H.-J.\ 2016, \prd,
    93, 044065
\bibitem[Giacomazzo \& Perna(2013)]{2013ApJ...771L..26G} Giacomazzo, B., \& Perna, R.\ 2013, \apjl,
    771, L26
\bibitem[Goldstein et al.(2017)]{2017ApJ...848L..14G} Goldstein, A., Veres, P., Burns, E., et al.\
    2017, \apjl, 848, L14
\bibitem[Hild et al.(2011)]{2011CQGra..28i4013H} Hild, S., Abernathy, M., Acernese, F., et al.\
    2011, Classical and Quantum Gravity, 28, 094013
\bibitem[Ho(2016)]{2016MNRAS.463..489H} Ho, W.~C.~G.\ 2016, \mnras, 463, 489
\bibitem[Hotokezaka et al.(2011)]{2011PhRvD..83l4008H} Hotokezaka, K., Kyutoku, K., Okawa, H.,
    Shibata, M., \& Kiuchi, K.\ 2011, \prd, 83, 124008
\bibitem[Hotokezaka et al.(2013)]{2013ApJ...778L..16H} Hotokezaka, K., Kyutoku, K., Tanaka, M., et
    al.\ 2013, \apjl, 778, L16
\bibitem[Johnson-McDaniel \& Owen(2013)]{2013PhRvD..88d4004J} Johnson-McDaniel, N.~K., \& Owen,
    B.~J.\ 2013, \prd, 88, 044004
\bibitem[Kiuchi et al.(2018)]{2018PhRvD..97l4039K} Kiuchi, K., Kyutoku, K., Sekiguchi, Y., \&
    Shibata, M.\ 2018, \prd, 97, 124039
\bibitem[Kiziltan et al.(2013)]{2013ApJ...778...66K} Kiziltan, B., Kottas, A., De Yoreo, M., \&
    Thorsett, S.~E.\ 2013, \apj, 778, 66
\bibitem[Lan et al.(2020)]{2020ApJ...890...99L} Lan, L., L{\"u}, H.-J., Rice, J., et al.\ 2020,
    \apj, 890, 99
\bibitem[L{\"u} \& Zhang(2014)]{2014ApJ...785...74L} L{\"u}, H.-J., \& Zhang, B.\ 2014, \apj, 785,
    74
\bibitem[L{\"u} et al.(2015)]{2015ApJ...805...89L} L{\"u}, H.-J., Zhang, B., Lei, W.-H., Li, Y., \&
    Lasky, P.~D.\ 2015, \apj, 805, 89
\bibitem[L{\"u} et al.(2017)]{2017ApJ...835..181L} L{\"u}, H.-J., Zhang, H.-M., Zhong, S.-Q., et
    al.\ 2017, \apj, 835, 181
\bibitem[L{\"u} et al.(2018)]{2018MNRAS.480.4402L} L{\"u}, H.-J., Zou, L., Lan, L., \& Liang,
    E.-W.\ 2018, \mnras, 480, 4402
\bibitem[Lasky \& Glampedakis(2016)]{2016MNRAS.458.1660L} Lasky, P.~D., \& Glampedakis, K.\ 2016,
    \mnras, 458, 1660
\bibitem[Lasky et al.(2014)]{2014PhRvD..89d7302L} Lasky, P.~D., Haskell, B., Ravi, V., Howell,
    E.~J., \& Coward, D.~M.\ 2014, \prd, 89, 047302
\bibitem[Lattimer \& Prakash(2004)]{2004Sci...304..536L} Lattimer, J.~M., \& Prakash, M.\ 2004,
    Science, 304, 536
\bibitem[Li et al.(2016)]{2016PhRvD..94h3010L} Li, A., Zhang, B., Zhang, N.-B., et al.\ 2016, \prd,
    94, 083010
\bibitem[Lin(2007)]{2007PhRvD..76h1502L} Lin, L.-M.\ 2007, \prd, 76, 081502
\bibitem[Lyford et al.(2003)]{2003ApJ...583..410L} Lyford, N.~D., Baumgarte, T.~W., \& Shapiro,
    S.~L.\ 2003, \apj, 583, 410
\bibitem[Markwardt(2009)]{2009ASPC..411..251M} Markwardt, C.~B.\ 2009, Astronomical Data Analysis
    Software and Systems XVIII, 411, 251
\bibitem[Metzger et al.(2011)]{2011MNRAS.413.2031M} Metzger, B.~D., Giannios, D., Thompson, T.~A.,
    Bucciantini, N., \& Quataert, E.\ 2011, \mnras, 413, 2031
\bibitem[Metzger et al.(2010)]{2010MNRAS.406.2650M} Metzger, B.~D., Mart{\'{\i}}nez-Pinedo, G.,
    Darbha, S., et al.\ 2010, \mnras, 406, 2650
\bibitem[Palenzuela et al.(2015)]{2015PhRvD..92d4045P} Palenzuela, C., Liebling, S.~L., Neilsen,
    D., et al.\ 2015, \prd, 92, 044045
\bibitem[Peng et al.(2019)]{2019ApJ...884L..34P} Peng, Z.-K., Yang, Y.-S., Shen, R.-F., et al.\
    2019, \apjl, 884, L34
\bibitem[Ravi \& Lasky(2014)]{2014MNRAS.441.2433R} Ravi, V., \& Lasky, P.~D.\ 2014, \mnras, 441,
    2433
\bibitem[Rezzolla et al.(2011)]{2011ApJ...732L...6R} Rezzolla, L., Giacomazzo, B., Baiotti, L., et
    al.\ 2011, \apjl, 732, L6
\bibitem[Rosswog et al.(2000)]{2000A&A...360..171R} Rosswog, S., Davies, M.~B., Thielemann, F.-K.,
    \& Piran, T.\ 2000, \aap, 360, 171
\bibitem[Rowlinson et al.(2010)]{2010MNRAS.409..531R} Rowlinson, A., O'Brien, P.~T., Tanvir, N.~R.,
    et al.\ 2010, \mnras, 409, 531
\bibitem[Shapiro et al.(1983)]{1983PhT....36j..89S} Shapiro, S.~L., Teukolsky, S.~A., \& Lightman,
    A.~P.\ 1983, Physics Today, 36, 89
\bibitem[Shapiro \& Teukolsky(1983)]{1983bhwd.book.....S} Shapiro, S.~L., \& Teukolsky, S.~A.\
    1983, Research supported by the National Science Foundation.~New York, Wiley-Interscience,
    1983, 663 p.,
\bibitem[Shibata \& Taniguchi(2006)]{2006PhRvD..73f4027S} Shibata, M., \& Taniguchi, K.\ 2006,
    \prd, 73, 064027
\bibitem[Sun et al.(2019)]{2019ApJ...886..129S} Sun, H., Li, Y., Zhang, B.-B., et al.\ 2019, \apj,
    886, 129
\bibitem[Valentim et al.(2011)]{2011MNRAS.414.1427V} Valentim, R., Rangel, E., \& Horvath, J.~E.\
    2011, \mnras, 414, 1427
\bibitem[Xiao et al.(2019)]{2019ApJ...879L...7X} Xiao, D., Zhang, B.-B., \& Dai, Z.-G.\ 2019,
    \apjl, 879, L7
\bibitem[Xiao \& Dai(2019)]{2019ApJ...878...62X} Xiao, D. \& Dai, Z.-G.\ 2019, \apj, 878, 62
\bibitem[Xue et al.(2019)]{2019Nature...568..198X} Xue, Y.-Q., Zheng, X.-C., Li, Y., et al. 2019,
    Nature, 568, 198
\bibitem[Yu et al.(2010)]{2010ApJ...715..477Y} Yu, Y.-W., Cheng, K.~S., \& Cao, X.-F.\ 2010, \apj,
    715, 477
\bibitem[Yu et al.(2013)]{2013ApJ...776L..40Y} Yu, Y.-W., Zhang, B., \& Gao, H.\ 2013, \apjl, 776,
    L40
\bibitem[Zhang et al.(2018)]{2018NatCo...9..447Z} Zhang, B.-B., Zhang, B., Sun, H., et al.\ 2018,
    Nature Communications, 9, 447
\bibitem[Zhang(2014)]{2014ApJ...780L..21Z} Zhang, B.\ 2014, \apjl, 780, L21
\bibitem[Zhang(2013)]{2013ApJ...763L..22Z} Zhang, B.\ 2013, \apjl, 763, L22
\bibitem[Zhang \& M{\'e}sz{\'a}ros(2001)]{2001ApJ...552L..35Z} Zhang, B., \& M{\'e}sz{\'a}ros, P.\
    2001, \apjl, 552, L35



\end{thebibliography}
\end{document}